\documentclass[aps,twocolumn,nofootinbib,preprintnumbers,prl,10pt]{revtex4-2}
\usepackage[utf8]{inputenc}
\usepackage[svgnames]{xcolor}
\usepackage[most]{tcolorbox}
\usepackage{caption}
\tcbset{colback=PaleGreen!0!white, colframe=NavyBlue, 
	highlight math style= {enhanced, 
		colframe=red,colback=red!10!white,boxsep=0pt}
}
\usepackage{relsize}

\usepackage{amsmath}
\usepackage[bottom]{footmisc}
\usepackage{braket}
\usepackage{graphicx}
\usepackage{mwe}
\usepackage[section]{placeins}
\usepackage{framed}
\usepackage{csquotes}
\usepackage{tikz}
\usetikzlibrary{decorations.markings}
\usetikzlibrary{decorations.pathmorphing}
\definecolor{shadecolor}{rgb}{0.90,0.90,0.90}
\usepackage{hyperref}
\usepackage{bbold}
\usepackage{subcaption}
\usepackage{pifont}
\usepackage{setspace}
\usepackage{amsmath, amssymb, amsthm, float, graphicx,amsfonts}
\usepackage{amsthm}
\usepackage{array, boldline, makecell, booktabs}

\theoremstyle{definition}

\hypersetup{colorlinks=true, linkcolor=DarkRed, citecolor=DarkRed,urlcolor=Indigo,linktocpage}
\def\beq{\begin{eqnarray}}\def\eeq{\end{eqnarray}}
\def\be{\begin{equation}}\def\ee{\end{equation}}
\def\bs{\begin{split}}\def\es{\end{split}}

\def\tz{{\tilde{z}}}
\def\wbeta{{\widehat\beta}}
\usepackage{bm}

\begin{document}
\title{\bf Dispersion relations, knot polynomials and the $q$-deformed harmonic oscillator}
\author{Aninda Sinha$^{a}$\footnote{asinha@iisc.ac.in} \\
\it ${^a}$Centre for High Energy Physics,
\it Indian Institute of Science,\\ \it C.V. Raman Avenue, Bangalore 560012, India. }

\begin{abstract}{We show that the crossing symmetric dispersion relation (CSDR) for 2-2 scattering leads to a fascinating connection with knot polynomials and q-deformed algebras. In particular, the dispersive kernel can be identified naturally in terms of the generating function for the Alexander polynomials corresponding to the torus knot $(2,2n+1)$ arising in knot theory. Certain linear combinations of the low energy expansion coefficients of the amplitude can be bounded in terms of knot invariants.  Pion S-matrix bootstrap data respects the analytic bounds so obtained. We correlate the $q$-deformed harmonic oscillator with the CSDR-knot picture. In particular, the scattering amplitude can be thought of as a $q$-averaged thermal two point function involving the $q$-deformed harmonic oscillator. The low temperature expansion coefficients are precisely the $q$-averaged Alexander knot polynomials.}
\end{abstract}
\maketitle

{\bf Introduction:} In the last few years, it has been realized that combining the power of dispersion relations with crossing symmetry leads to two-sided bounds on the low energy Taylor expansion coefficients (Wilson coefficients) of 2-2 scattering amplitudes \cite{nima1, nima2, tolley1, sch1, snowmass}. There are two complementary dispersion relations in the literature that have been used in these studies--(a) The fixed-$t$ dispersion relation where crossing symmetry is not manifest and (b) The crossing symmetric dispersion relation (CSDR) where crossing symmetry is manifest but locality is not \cite{AK, ASAZ, GSZ}. The CSDR has led to a surprising connection with an area of mathematics \cite{rogo, branges, duren} called Geometric Function Theory (GFT) \cite{HSZ, RS,AZ}.

In this paper, we will find a very interesting connection between the CSDR, knot polynomials and $q$-deformed algebras.   It has been known for a while that there are surprising connections between knot theory and areas of physics like quantum field theory and statistical physics \cite{kauffman, baez, wittenjones, wittenpop}. In the context of perturbative Feynman diagrams in QFT, a connection with knot theory has been suggested in the literature and reviewed in \cite{kreimer}. In particular, it was shown in \cite{kreimer} by considering specific examples in $\phi^4$ perturbation theory, that torus knots play a role in such calculations. 
One of the main goals of this paper is to use a non-perturbative dispersive representation to show how a class of knot polynomials arise naturally in the description of 2-2 scattering of identical particles. Although our discussion will focus on 2-2 scattering of scalars, as was shown in \cite{spinning}, the CSDR form used in this paper is applicable to the situation where we have external spinning particles as well. Quite surprisingly, as we will elaborate later, the form of the dispersive representation is ubiquitous and arises in CFT \cite{GSZ, bs, liendo}. Hence knot polynomials will play a role in discussing conformal correlators as well. At the onset, we should emphasise that our analysis will only suggest a possible role of knots in the description of scattering. We do not pretend to have a rigorous justification for the same. We will nevertheless attempt to interpret the knot parameter arising in our description in terms of q-deformed algebras.

 The oldest of the knot polynomials is the Alexander polynomial introduced by J. W. Alexander in 1923 \cite{jw}. This will be the main character in the present paper. We will show how this polynomial naturally appears in the CSDR. One of the simplest quantum field theories we study in an introductory course in QFT is the $\phi^3$ theory and its cousins. In the crossing symmetric variable, the Alexander polynomials for a certain class of knots, called the torus knots, make a natural appearance in the discussion of 2-2 scattering in such theories. The kernel in the CSDR resembles a linear combination of $\phi^3$ and $\phi^4$ theories with the mass parameter integrated over in a certain way. The CSDR gives a non-perturbative represenation of 2-2 scattering describing, for instance, pion scattering.
 
\begin{figure}[hbt]
		\includegraphics[width=0.4\textwidth]{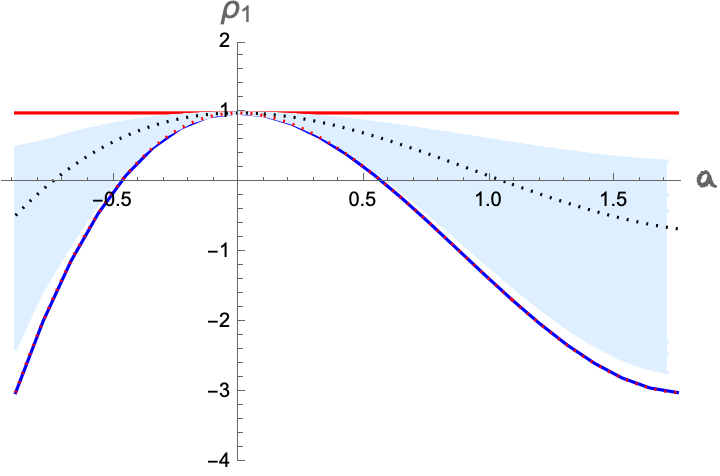}
		\label{fig:n1}
\vskip 0.5cm
		\includegraphics[width=0.4\textwidth]{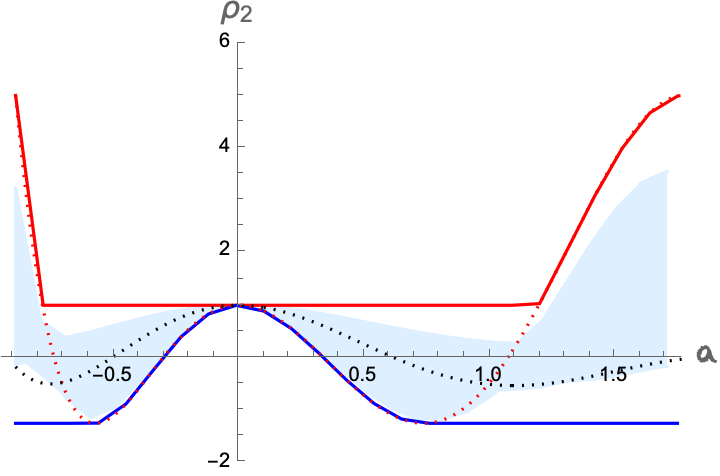}
		\label{fig:n2}
	\caption{$\rho_n$ vs $a$. The blue shaded region is the spread of values obtained from the pion S-matrix bootstrap \cite{andrea, ABPHAS, pionsupp1}, with $m_\pi=1$. The red dotted line is tree level $\phi^3$ with $m=1$ while the black dotted  line is one loop $\phi^4$ with $m=1$. The solid blue line is the anaytic minimum and the solid red line is the analytic maximum given by eq.(\ref{maineq}). For $\rho_1$ the red dotted line and the theoretical minimum are on top of each other.}
	\label{fig:n1n2}
\end{figure}

Some of the two-sided bounds, alluded to above, can be recast in terms of the knot invariants involving the Alexander polynomial of a $(2,2n+1)$ torus knot evaluated at a special value for the polynomial variable. This provides a novel geometric picture for understanding these bounds. As we will show using the CSDR, the derivatives of the scattering amplitude with respect to the crossing symmetric variable can be thought of as an average over the Alexander polynomials. Eq.(\ref{maineq}) and Eq.(\ref{cn}) give bounds on the derivatives of the amplitude, and hence on Wilson coefficients, in terms of the Alexander polynomials; eq.(\ref{cn}) gives the global maxima for the absolute values in terms of the knot crossing number. We use the pion S-matrix bootstrap to test these bounds (fig.\ref{fig:n1n2}) and find convincing evidence. Similar bounds exist for a wide variety of cases, e.g. in CFT and scattering of external spinning particles as we will discuss below. 

Since the dispersion relation involves a kind of averaging over the knot parameter, it begs the question what physical picture can we keep in mind while thinking about the CSDR-knot connection. To tackle this question, we observe that the generating function of the Alexander polynomial makes a mysterious appearance in another context in physics. When we consider the $q$-deformed harmonic oscillator as introduced by Biedenharn \cite{biedenharn} and Macfarlane \cite{macfarlane}, then it turns out that the thermal two point function involving the deformed oscillators \cite{qgroup} is precisely the generating function of the Alexander polynomial! This connection does not appear to have been pointed out in the literature so far.  We will make some preliminary observations what this connection can teach about scattering. In particular, we derive the map in eq.(\ref{map1}) which relates the CDSR to a $q$-averaged thermal two point function of the $q$-deformed oscillator, and use it to explain certain observations about 2-2 scattering of identical particles in specific limits. We will begin by reviewing some key points about the Alexander polynomials.\\

{\bf Some knot theory:} The key player in our discussion will be the Alexander polynomials for  the $(2,2n+1)$ torus knots. In the appendix, we review torus knots in more detail. For now, it is sufficient to note that the $(2,2n+1)$ torus knot is one that wraps the longitude of the torus twice and the meridian $2n+1$ times. These knots are distinguishable using the Alexander knot polynomials. For the $(2,2n+1)$-torus knots these are given by
\begin{equation}\label{torusA}
{\mathcal A}^{(2,2n+1)}(q)= q^{-n}\frac{q^{2n+1}+1}{q+1}\,.
\end{equation}
While it is not obvious that these are Laurent polynomials in $q$, one can easily  explicitly verify that they are. $(2n+1)$ is also the number of crossings for the torus knot and is an invariant. Furthermore, we note that ${\mathcal A}^{(2,2n+1)}(1)=1$ while $|{\mathcal A}^{(2,2n+1)}(-1)|=2n+1$, the latter is called the knot determinant and is a knot invariant. In the situation where $q$ is a pure phase, it can be verified that $2n+1$ is the maximum absolute value for the knot polynomial.

{\it Key observation}:
The key mathematical insight that enables us to correlate the CSDR with the Alexander polynomials is the relation of the latter with the Chebyshev polynomials of the second kind. This observation was made in \cite{gavrilik1, gavrilik2}.  Denote by $U_n(x)$ the Chebyshev polynomials of the second kind such that $U_0(x)=1, U_1(x)=2x, U_2(x)=4x^2-1$. The generating function of these polynomials is given by
\begin{equation}
\frac{1}{1-2x z+z^2}=\sum_{n=0}^\infty z^n U_n(x)\,.
\end{equation}
This relation will be the key in relating the CSDR to the Alexander knot polynomials. The relation between the $U_n(x)$ and ${\mathcal A}^{(2,2n+1)}(t)$ is simply the following:
\begin{equation}\label{key1}
{\mathcal A}^{(2,2n+1)}(q)=U_n(x)-U_{n-1}(x)\,, \quad {\rm where}\quad 2x=q+\frac{1}{q}\,.
\end{equation}
Equivalently we can verify \cite{ffoot}
\begin{equation}\label{key2}
\frac{1-z}{1-2x z+z^2}=\sum_{n=0}^\infty z^n {\mathcal A}^{(2,2n+1)}(q)\,.
\end{equation}
with the above relation in eq.(\ref{key1}) between $q$ and $x$.
Generalizations of eq.(\ref{key1}) for the torus knots of the type $(3,p)$ can be found in \cite{gavrilik2}. We will touch upon these generalizations below.\\

{\bf Torus knots in QFT}:
Let us begin by discussing 2-2 scattering of $\phi\phi\rightarrow \phi \phi$ in tree level $\phi^2 \psi$ theory\cite{foot1}, one of the simplest examples one encounters in a first course in QFT. We will treat $\phi$ to be a massless scalar and $\psi$ to be massive scalar of mass $m$.  The amplitude, up to an overall coupling constant, is given by
\begin{equation}
{\mathcal M}(s,t)=\frac{1}{m^2-s_1}+\frac{1}{m^2-s_2}+\frac{1}{m^2-s_3}\,,
\end{equation}
subject to the constraint $s_1+s_2+s_3=0$.  Consider the following change of variables \cite{AK}:
\begin{equation}
s_k=a \left(1-\frac{(z-z_k)^3}{z^3-1}\right)\,.
\end{equation}
Here $z_k= \exp(2 \pi i (k-1)/3)$ are the cube-roots of unity. One can check that $s_1+s_2+s_3=0$ and that 
\begin{equation}
a= \frac{s_1 s_2 s_3}{s_1 s_2+s_1 s_3+s_2 s_3}\,.
\end{equation}
The effect of this transformation is to map branch cuts in the $s_1,s_2,s_3$ planes to arcs of a unit circle \cite{AK, ASAZ}. 
It is easy to check that the amplitude is given by \cite{foot2}
\begin{eqnarray}\label{phi2psi}
{\mathcal M}(\tz,a)-\frac{3}{m^2}&=&\frac{\wbeta(\frac{a}{m^2})}{m^2}\frac{\tilde z}{1-2\xi \tilde z+\tilde z^2}, \label{simpgen}
\end{eqnarray}
where  $\tilde z\equiv z^3$ and $\wbeta,\xi$ are given by
\begin{equation}\label{xidef}
\wbeta(\alpha)=27\alpha^2(3\alpha-2)\,,\quad 2\xi=2-27 (\frac{a}{m^2})^2+27 (\frac{a}{m^2})^3\,.
\end{equation}
Thus using eq.(\ref{key2}) we see that $(1-\tz)({\mathcal M}(\tz,a)-\frac{3}{m^2})$ is the generating function of the $(2,2n+1)$-Alexander knot polynomials with knot parameter $q$ given via $2\xi=q+1/q$. We can re-express individual knot polynomials in terms of Taylor expansion coefficients of the amplitude. Define:
\be\label{rhodef}
\rho_n=\frac{1}{\partial_\tz {\mathcal M}(\tz,a)}\left (\frac{ \partial_\tz^{n+1} \mathcal M(\tilde z,a)}{(n+1)!}-\frac{ \partial_\tz^{n}  \mathcal M(\tilde z,a)}{n!}\right ){\bigg |}_{\tz=0}\,.
\ee
Now using eq.(\ref{key1}), we easily find
\begin{eqnarray}
\rho_n= U_n(\xi)-U_{n-1}(\xi)= {\mathcal A}^{(2,2n+1)}(q)\,.
\end{eqnarray}
This relates the derivatives of the scattering amplitude to the Alexander polynomials of $(2,2n+1)$ torus knots. Note that $\xi=1$ (equivalently $q=1$) corresponds to $a/m^2=0,1$ while $\xi=-1$ (equivalently $q=-1$) gives $a/m^2=-1/3$ or $2/3$. In order to avoid singularities inside the unit disc $|\tilde z|=1$, we need $-1/3\leq a/m^2\leq 2/3$. So the crossing number or knot determinant emerges at the boundary values of this domain. This feature will carry over to the CSDR. It is indeed gratifying to note that there is a connection between the oldest knot polynomial and one of the simplest quantum field theories. Now we turn to a non-perturbative representation of scattering amplitudes via the CSDR and examine what role knot polynomials play there.

{\bf CSDR and bounds with ${\mathcal A}^{(2,2n+1)}$: }
The idea behind a CSDR is to write a dispersion relation treating $a$ as a parameter and fixed and the dispersive variable to be $z$. As reviewed in \cite{ASAZ}, a fully crossing symmetric amplitude then is a function of $a, z^3$. 
In \cite{RS}, it was shown that the CSDR in \cite{AK, ASAZ} can be written in the form of the Robertson representation of typically real functions in the context of GFT\cite{ffoot3}. This was then instrumental in giving two sided bounds on the (ratios of) Wilson coefficients of the amplitude. In general, in order to describe scattering of identical particles\cite{foot3} in a situation where in the complex-$s_1$ plane, there is a gap between the $s_1$-channel and $s_3$-channel cuts, and where the amplitude for large $|s_1|$ falls off faster than $|s_1|^2$, we have:
\begin{eqnarray}\label{disp}
{\mathcal M}(\tilde z,a)&=&\alpha_0-\frac{2\mathcal{N}}{\pi} \int_{\xi_{min}}^{\xi_{max}} \, d\mu(\xi) \frac{\tilde z}{\tilde z^2-2\xi \tilde z+1} 
\end{eqnarray}
\begin{equation}
{\mathcal N}=-\frac{\pi}{2}\partial_\tz {\mathcal M}(\tz,a)|_{\tz=0}\,,
\end{equation}
and
\begin{equation}
d\mu(\xi)={\mathcal N}^{-1} d\xi Im {\mathcal M}(s_1'(\xi,a),s_2(\xi,a))
\end{equation}
defines a measure
with $Im \mathcal{M}(s_1,s_2)$ denoting the $s_1$-channel discontinuity and $s_2=-\frac{s_1'}{2}(1-(\frac{s_1'+3a}{s_1'-a})^{1/2})$. More details can be found in \cite{ASAZ}. Here $2\xi=2-27(a/s_1')^2+27(a/s_1')^3$ as in eq.(\ref{xidef}).  So integral over $\xi$ is an integral over $s_1'$ since $a$ is fixed.
We will choose the cut in the $s_1$ plane to begin at $s_1=8/3$ and this sets the normalization for us. Our focus will be on the interval $-8/9\leq a\leq 16/9$ as in this range the amplitude is typically real \cite{RS}; this will enable us to compare with the Bieberbach conjecture. In this range $\mathcal{N}$, $d\xi/ds_1'$ and $Im \,\mathcal{M}$ are all positive \cite{HSZ, RS} so $\mu(\xi)$ can be thought of as a probability measure. It is easy to check that $\xi_{max}=1$ while $\xi_{min}=1-\frac{243 a^2}{128}+\frac{729 a^3}{1024}$. When $a=-8/9,16/9$, $\xi_{min}=-1$ else it is $>-1$. 
%
We will now derive an interesting inequality.
Starting with eq.(\ref{rhodef}) and eq.(\ref{disp}) we have
\begin{equation}
 \rho_n=\int_{\xi_{min}}^{\xi_{max}} \, d\mu(\xi){\mathcal A}^{(2,2n+1)}(q)\,,
 \end{equation}
 where $2\xi=q+\frac{1}{q}$. Using the positivity of the measure and $\int_{\xi_{min}}^{\xi_{max}}d\mu(\xi)=1$, we easily find
 \begin{eqnarray}
{\mathcal A}^{(2,2n+1)}(q){\bigg|}_{min} \leq \rho_n\leq {\mathcal A}^{(2,2n+1)}(q){\bigg|}_{max} \label{maineq}\,.
\end{eqnarray}
%
%
Here the maximum and minimum values of the knot polynomials are for the argument in the range $\xi_{min}\leq \xi\leq \xi_{max}$. A few examples when $-1\leq \xi\leq 1$ are: $-3\leq \rho_1\leq 1, -1.25\leq \rho_2\leq 5, -7\leq \rho_3\leq 1.63$. These results do not follow from the extrema of $U_n$'s which are extremized at different values of $\xi$ compared to ${\mathcal A}^{(2,2n+1)}$. The absolute value of the extremum satisfies
\begin{equation}\label{cn}
|\rho_n|\leq {\bigg |}{\mathcal A}^{(2,2n+1)}(-1){\bigg |}={\rm crossing~number}\,,
\end{equation}
where we have used that the knot determinant/crossing number maximizes $|{\mathcal A}^{(2,2n+1)}(q)|$ for any $a\in[-8/9,16/9]$. This gives a knot theory interpretation of the global maximum of $|\rho_n|$. The equality occurs for theories like the tree-level $\phi^3$ theory. Near $a\sim 0$, $\xi\sim 1$ and since ${\bigg |}{\mathcal A}^{(2,2n+1)}(1){\bigg |}=1$, we find that $\rho_n\sim 1$ near $a\sim 0$. Plots of $\rho_1$ and $\rho_2$ vs $a$ for the pion bootstrap are shown in fig.(\ref{fig:n1n2}). The data was obtained from the S-matrix bootstrap for pions ($\pi^0\pi^0\rightarrow \pi^0\pi^0$) in \cite{ABPHAS} following \cite{andrea}.  In the pion bootstrap, one assumes unitarity, crossing symmetry and the input of the $\rho$-resonance mass in the complex $s$-plane \cite{pionsupp1}. In this manner, we get a family of consistent S-matrices. Since tree-level $\phi^3$ theory saturates the lower bound for $\rho_1$ in eq.(\ref{maineq}), this is as tight as possible and imposing non-linear unitarity will not make the $\rho_1$ lower bound any tighter. S-matrix bootstrap data respects the analytic bounds very well.

At this stage, we should point out that $\rho_n$ was defined such that we get bounds in terms of the Alexander polynomials. Since ${\mathcal M}(\tz,a)$ and $(1-\tz){\mathcal M}(\tz,a)$ have the same information in so far as the Taylor expansion coefficients of the amplitude are concerned, the pictures in terms of the Chebyshev and Alexander polynomials are equivalent. For instance, as shown in the appendix, the global bounds can be rederived using different techniques which make use of the Chebyshev polynomials. Nevertheless, only the knot picture enables us to interpret the bounds in terms of interesting topological quantities like the crossing number or knot determinants.  Further, what we find fascinating is that the Alexander polynomials for the $(2,2n+1)$-torus knots appear to provide a complete basis for the amplitude. In order to probe this further, it is important to have a physical picture as to what the averaging in eq.(\ref{disp}) over the knot parameter $q$ means.

{\bf $q$-deformed harmonic oscillator:}
Eq.(\ref{disp}) is telling us to think of the amplitude as a certain integral over the mass parameter for $\phi^3$ theories (with a specific $\phi^4$ contact term). In terms of knots, the Taylor expansion coefficients of the amplitudes involve an ``averaging'' over the knot parameter. What physical picture can we keep in mind while trying to interpret this? To address this, we will now correlate the CSDR with the $q$-deformed oscillator \cite{qgroup} which features in the discussion of the quantum group $SU_q(2)$. The $q$-deformed oscillator was introduced independently by Biedenharn and Macfarlane in 1989 and has been studied in great detail in the literature. The book \cite{qgroup} is a good reference for the material used below. The $q$-oscillator uses 3 generators $b,b^\dagger, N$ satisfying $b b^\dagger-q b^\dagger b=q^{-N}, [N,b]=-b, [N,b^\dagger]=b^\dagger$. We have $(b^\dagger)^\dagger=b, N^\dagger=N$. Here $q$ is either a real number or  a complex number with unit modulus. In the latter case, which will be our focus, we also have $b b^\dagger-q^{-1} b^\dagger b=q^N$ so that 
\be
b^\dagger b=[N]_q\,, \qquad b b^\dagger=[N+1]_q\,,
\ee
where we have introduced the $q$-number
\be
[x]_q\equiv \frac{q^x-q^{-x}}{q-q^{-1}}\,.
\ee
When $x$ is an integer, it can be checked that the $q$-numbers are just Chebyshev polynomials of the second kind \cite{gavrilik1, gavrilik2}. Specifically,
\be
[n]_q=U_{n-1}\left(\frac{1}{2}(q+\frac{1}{q})\right)\,.
\ee
When $q\rightarrow 1$, $[x]_q\rightarrow x$. For a phase $q=e^{i\theta}$, we have $[n]_q= \sin(n\theta)/\sin(\theta)$ which obeys $-\kappa_n\leq [n]_q\leq n$ with $-\kappa_n={\rm min}_{0\leq \theta\leq 2\pi} \sin(n\theta)/\sin(\theta)$.  This is what leads to the Bieberbach-Rogosinski bounds in GFT \cite{RS}. The Fock space states $|n\rangle$ obey $N|n\rangle=n |n\rangle, b^\dagger |n\rangle=\sqrt{[n+1]_q}|n+1\rangle, b|n\rangle=\sqrt{[n]_q}|n-1\rangle$ making it clear that $[n]_q$ has to be finite. 
For the $q$-oscillator, we can define the Hamiltonian as \cite{qgroup}
\be
H=w N\,.
\ee
Quite remarkably, the thermal expectation value of $b^\dagger b$ is given by \cite{qgroup}
\be\label{aadag}
\langle b^\dagger b \rangle_{\beta,q}=
\frac{\tz(1-\tz)}{\tz^2-2\xi \tz+1}\,,
\ee
where $ \tz\equiv e^{-\beta w}$ and $2\xi=q+1/q$.
This is precisely of the form of the generating function (see eq.(\ref{simpgen})) for the Alexander polynomials in terms of the dispersive kernel. The expansion around $\tz=0$ corresponds to zero temperature limit $\beta\rightarrow \infty$. 
Using this, we can write the map
\be\label{map1}
 {\mathcal M}(\tilde z,a)=\alpha_0 -\frac{2\mathcal{N Z}}{\pi}  \int_{\xi_{min}}^{\xi_{max}} \, d\mu(\xi) \, \langle b^\dagger b \rangle_{\beta,q} \,, \mathcal{Z}=\frac{e^{\beta w}}{e^{\beta w}-1}\,,
\ee
where $2\xi=q+1/q$ as before and $\mathcal{Z}$ is the thermal partition function for the $q$-oscillator. When $|\xi|\leq 1$, we have $q$ to be complex and $|q|=1$. This is precisely the case for $q$ discussed above! In other words, the amplitude is related to the $q$-average of the expectation value of $b^\dagger b$ with inverse temperature set by $-(\ln \tz)/w$. The measure factor involves the partial wave amplitudes and hides the dynamical information. In the forward limit when $a\rightarrow 0$, as well as for large values of the dispersive variable $s_1'$ we have $q \rightarrow 1$. So in these regimes, we have the usual undeformed oscillator picture where the corresponding knot polynomial is unity. 

This last observation provides a physical interpretation of the following feature of amplitudes. Near $a\sim 0$ (also for $s_1'\gg a$), all coefficients of $\tz^n$ are always negative. This is because in this region we have an approximate description in terms of the usual simple harmonic oscillator. According to eq.(\ref{map1}), the $\tz^n$ coefficients are related to the expectation value of the usual number operator which is positive. Explicitly $[n]_q\rightarrow n>0$ and because ${\mathcal N}>0$ in eq.(\ref{map1}), all coefficients are negative. Due to the map, it is now also clear why the coefficients of $\tz^n$ are bounded; this is an inherited property from the construction whereby $[n]_q$ is bounded when $q$ is a phase. The two sided bounds on Wilson coefficents arise due to the two sided bounds on the $q$-numbers or equivalently due to the finite norms of the Fock space states. Explicitly, with $\mu(\xi)$ a probabilistic measure factor, we have the obvious inequality $-\kappa_n\leq \int d\mu(\xi) [n]_q \leq n$ which is what was used using different methods relying on GFT in \cite{RS}.

In summary, the $q$-deformed harmonic oscillator map provides an equivalent description of the CSDR.  When $q$ is a phase, the measure factor in the CSDR is positive and is a probability measure.  Since each value of $q$ corresponds to a specific theory, we are essentially averaging over theories. \\

{\bf Applications in CFT:} Our discussion so far can be easily extended to CFTs. Consider the four point correlation function of identical scalar primary operators. In \cite{GSZ}, the dispersive representation of Mellin amplitudes for this case was considered and found to be identical in form as the QFT case. In \cite{bs}, it was recently shown that the position space dispersion relation which makes the symmetry under the cross-ratio interchange manifest has exactly the same kernel in the $\tilde z$ variable! Hence, whatever we have discussed so far readily carries over to the situations where the CSDR is applicable. For instance eq.(\ref{maineq}) and eq.(\ref{cn}) will be applicable in their present forms in these cases. The only additional analysis needed is to determine the range of parameters where the measure is positive. In fact, even in situations where we need higher subtractions, in \cite{RS} it was shown that the form of the kernel factorizes into what we have been using so far, times polynomials in $\tilde z/(1-\tilde z)^2$. Thus with minimal changes, the discussion in this paper readily carries forward to those situations as well. What is perhaps even more surprising is that in the case where we have defect CFTs, e.g. \cite{liendo}, the dispersion relation there also has the same kernel as what we have considered here. The discussion in this paper will carry over to all these diverse situations.
\vskip 0.5cm
{\bf Discussion: }
 We derived bounds on the combination of derivatives of the amplitude in terms of knot invariants. The expression in eq.(\ref{maineq}) was in the form of an average over knot polynomials. We gave a physical interpretation of this quantity in terms of the $q$-deformed oscillator. In the future, it will be useful to develop the $q$-deformed oscillator picture further. An example of a question that would be interesting to answer is: In terms of the $q$-oscillator picture, what restrictions on the averaging correspond to local theories? We examine this question briefly using a string theory example in the appendix.
 
 Note that eq.(\ref{maineq}) holds for any regular typically real function inside the unit disc which is known to respect the Robertson representation. Thus, this connects typically real functions with knot theory--a connection that has not been pointed out previously, to the best of our knowledge.
  One can ask if other knot polynomials like the Jones polynomial can be expressed in terms of the amplitude using the CSDR. It can be shown  that in this case, the Jones polynomials for the torus $(2,2n+1)$ is a linear combination of derivatives including up to $3n+1$ terms instead of the two terms for the Alexander case. Thus in a very tangible sense, the Jones polynomials are ``more complex'' from the Amplitudes perspective. It will be interesting to investigate this kind of ``complexity'' in more generality using amplitudes.  It will also be interesting to examine the knot theory--perturbative Feynman diagram connection in the program of \cite{kreimer} using the CSDR.\\
  \vskip 0.5cm
{\bf Acknowledgments:} 
I thank B. Ananthanarayan, Agnese Bissi, Rajesh Gopakumar, Parthiv Haldar, P. Ramadevi, Prashanth Raman and Ahmadullah Zahed for useful discussions.  I acknowledge support from MHRD, Govt. of India, through a SPARC grant P315 and DST through the SERB core grant CRG/2021/000873. Hospitality of IAS, Princeton is gratefully acknowledged where v1 of this work was presented.


\section{appendix}

\section {\bf Alexander polynomials in knot theory: }
\begin{figure}[!htb]
	\hskip 1cm
	\begin{subfigure}[b]{0.25\textwidth}
		\centering
		\includegraphics[width=\textwidth]{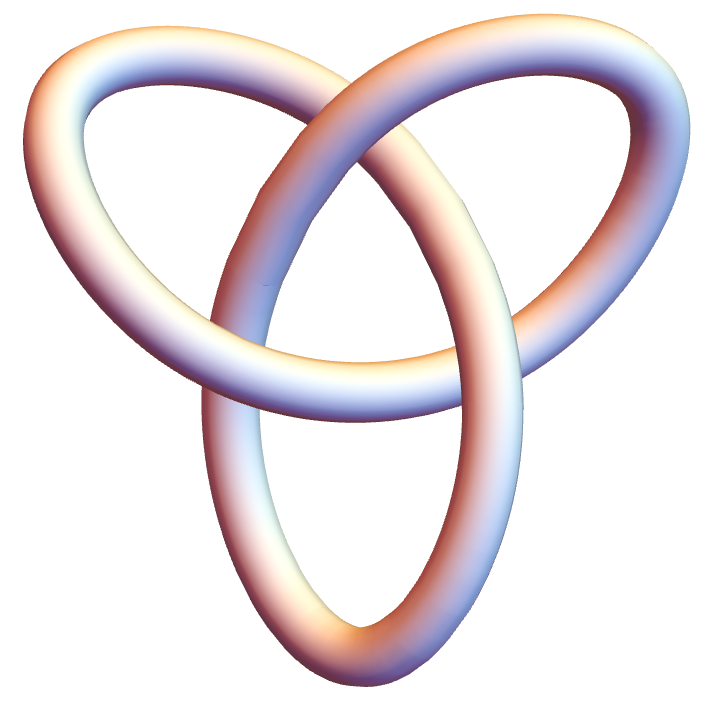}
		\caption{Trefoil (2,3)}
		\label{fig:trefoil}
	\end{subfigure}
	\hfill
	\begin{subfigure}[b]{0.25\textwidth}
		\hskip -.5cm
		\includegraphics[width=\textwidth]{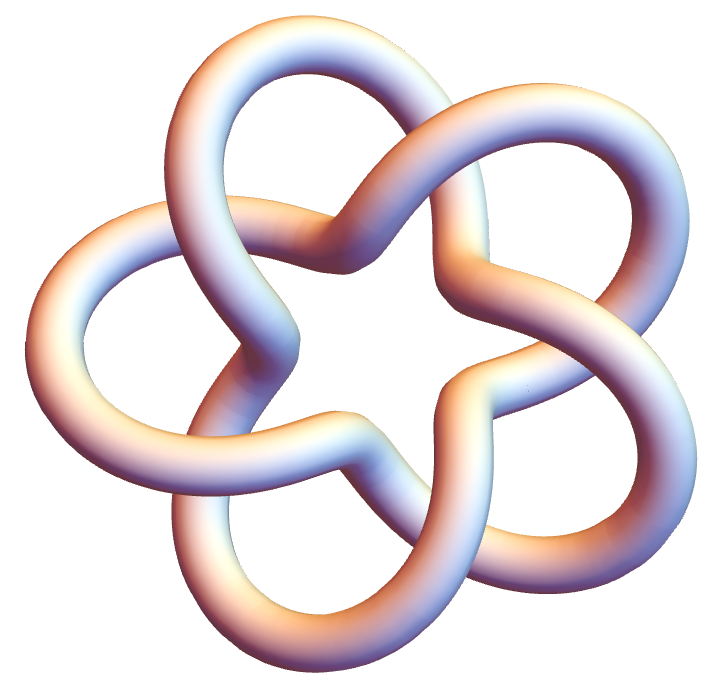}
		\caption{Pentafoil (2,5)}
		\label{fig:solomon}
	\end{subfigure}
	\hfill
	\begin{subfigure}[b]{0.25\textwidth}
		\hskip -0.5cm
		\includegraphics[width=\textwidth]{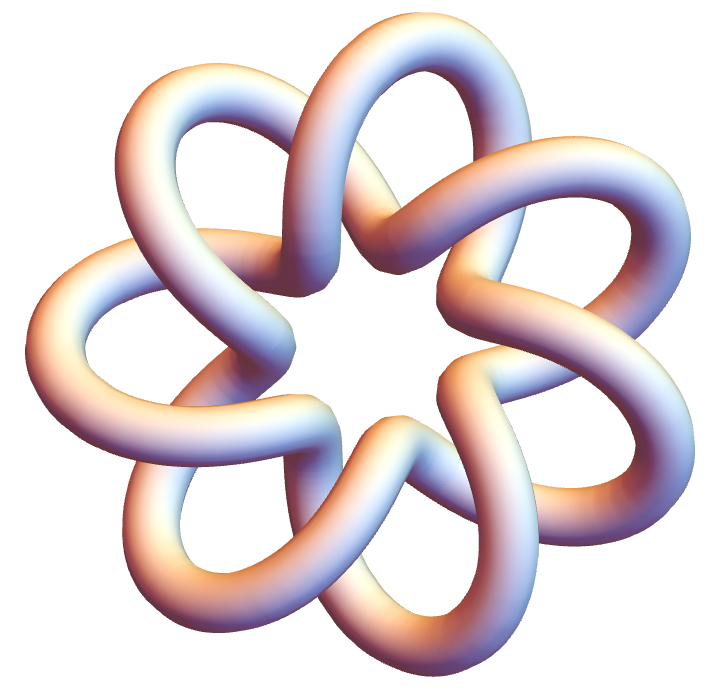}
		\caption{Heptafoil (2,7)}
		\label{fig:27}
	\end{subfigure}
	\caption{Some torus knots. Figures generated in mathematica.}
	\label{fig:knots}
\end{figure}

In this section, I will briefly review some background material on Alexander polynomials for torus knots \cite{wiki}.
Let me begin with a lightning review of knot polynomials in general. Knot polynomials allow us to characterise knots. Two knots are equivalent if they can be transformed into each other using a finite number of Reidemeister moves \cite{reide}. If the knot polynomials of two knots are different, then the knots are not equivalent. The converse is not true; two different knots can have the same knot polynomial. Several famous knot polynomials are known--these include Alexander, Conway-Alexander, Jones, Kauffmann and HOMFLY-PT. The oldest and perhaps the simplest of these is the Alexander polynomials. We will see that in the CSDR, Alexander polynomials make a natural appearance, so our focus will be on this. Torus knots are labeled by two co-prime integers $(p,q)$. Here the curve depicting the knot on the torus traverses $p$ times along longitude and $q$ times along meridian.  For $(p,q)$ co-prime, the torus knot is prime; in other words, it cannot be decomposed into smaller knots, much like how we define prime numbers. In the CSDR, we will see that the torus knot $(2,2n+1)$ where $n$ is a positive integer, makes an appearance. For a $(p,q)$-torus knot, the Alexander polynomials are given by \footnote{Since it is conventional to use $(p,q)$ to label prime knots, we will use $t$ as the knot parameter in the appendix.}
\begin{equation}
{\mathcal A}^{(p,q)}(t)=t^{-g} \frac{(t^{pq}-1)(t-1)}{(t^p-1)(t^q-1)}\,,
\end{equation}
where the genus-$g$ of the knot is given by
\begin{equation}
g=\frac{1}{2}(p-1)(q-1)\,.
\end{equation}
In the case of interest $p=2, q=2n+1$ so that the genus is $n$ and 
\begin{equation}\label{torusA}
{\mathcal A}^{(2,2n+1)}(t)= t^{-n}\frac{t^{2n+1}+1}{t+1}\,.
\end{equation}
 Further the crossing number of the $(2,2n+1)$ knot is $c=2n+1$ and is also a topological invariant. In figure. (\ref{fig:knots}) we show the knots corresponding to $n=1,2,3$. 

The Alexander polynomials for the $n=1,2,3$ cases are shown below:
\begin{eqnarray}
{\mathcal A}^{(2,3)}(t)&=& t+\frac{1}{t}-1\,,\\
{\mathcal A}^{(2,5)}(t)&=& t^2+\frac{1}{t^2}-t-\frac{1}{t}+1\,,\\
{\mathcal A}^{(2,7)}(t)&=& t^3+\frac{1}{t^3}-t^2-\frac{1}{t^2}+t+\frac{1}{t}-1\,.
\end{eqnarray}
The Alexander, Conway, Jones polynomials for the pentafoil knot are the same as the knot\footnote{The nomenclature means that the (prime) knot has 10 crossings and in some standard list is the 132nd knot with 10 crossings.} $10_{132}$ which is distinguished by the Kauffman polynomial.
The Alexander polynomials for any knot satisfy two important properties that we note:
\begin{eqnarray}
{\mathcal A}(1)&=& 1\,,\\
|{\mathcal A}(-1)| &=&{\rm invariant}\,.
\end{eqnarray}
The quantity $|{\mathcal A}(-1)|$ is called the knot determinant and is a special knot invariant. For the torus knot $(2,2n+1)$ we have ${\mathcal A}^{(2,2n+1)}(1)=1$ and
\begin{equation}
|{\mathcal A}^{(2,2n+1)}(-1)|=2n+1=c\,,
\end{equation}
which is the same as the crossing number
and is the maximum absolute value for the Alexander polynomial in the interval $\frac{1}{2}(t+\frac{1}{t})\in [-1,1]$. In the main text we will switch to the notation $t\rightarrow q$.

\section {\bf Deriving global bounds on $|\rho_n|$ using GFT:}
We can easily rederive the global bounds for $\rho_1$ and $\rho_2$ in the main text using the Bieberbach-Rogosinski bounds for a typically real function \cite{foot6} which is regular inside the unit disc. Writing ${\mathcal M}(\tz,a)$ as
\begin{equation}
{\mathcal M}(\tz,a)=\alpha_0+\sum_{n=1}^\infty \alpha_n \tz^n\,,
\end{equation}
we know that it is a typically real function inside the unit disc $|\tz|<1$ for $a\in [-8/9,16/9]$ and obeys the Bieberbach-Rogosinski inequalities \cite{RS}
\begin{equation}\label{BR}
-\kappa_n\leq \frac{\alpha_n}{\alpha_1}\leq n\,,\quad n\geq 2
\end{equation}
where $\kappa_n$ is $n$ for even $n$ and is some number less than $n$ for odd $n$. The precise number is unimportant. As an example, consider $n=2$. Then the lhs of eq.(15) in the main text is $|\alpha_2/\alpha_1-1|$. Using the lower bound on $\alpha_2/\alpha_1$, which is -2, we find $|\alpha_2/\alpha_1-1|\leq 3$. Similarly for $n=3$ we have $|\alpha_3/\alpha_1-\alpha_2/\alpha_1|$. Here we need the upper bound for $\alpha_3/\alpha_1$ and lower bound for $\alpha_2/\alpha_1$. Together this yields $|\alpha_3/\alpha_1-\alpha_2/\alpha_1|\leq 5$.  It is easy to check that the generalization of this argument leads to 
\begin{equation}
{\bigg |}\frac{\alpha_n}{\alpha_1}-\frac{\alpha_{n-1}}{\alpha_1}{\bigg |}\leq 2n+1\,,
\end{equation}
using eq.(\ref{BR}). The rhs is precisely the knot determinant for the torus knot $(2,2n+1)$. Note however,  that the present analysis gives a stronger bound on $\rho_n$ in terms of the knot polynomial, which is $a$-dependent unlike the $a$-independent eq.(\ref{BR}). Writing ${\mathcal M}(s_1,s_2)=\sum {\mathcal M}_{pq}x^p y^q$ with $x=-(s_1 s_2+s_1 s_3+s_2 s_3), y=-s_1 s_2 s_3$ and defining $w_{pq}={\mathcal W}_{pq}/{\mathcal W}_{10}$, we get the same two-sided bounds on $w_{01}$ as GFT \footnote{This is the same that arises from the numerical techniques of \cite{sch1}.} and somewhat stronger results for $w_{11}, w_{20}, w_{02}$ than using eq.({\ref{BR}) directly.

\section {\bf Some details of the pion bootstrap:}
Here we will summarize the key assumptions used in obtaining the pion S-matrices using numerical S-matrix bootstrap following \cite{andrea, ABPHAS}. To obtain pion S-matrices, we assume \cite{andrea} $O(3)$ symmetry, unitarity and crossing symmetry. In addition to these, we restrict the S-matrices by inputting the location of the $\rho$ resonance. In the space of the Adler zeros, we consider three different regions which are obtained as follows ($m_\pi=1$):

\begin{enumerate}
\item Pion lake: This is obtained using the assumptions stated above and was first obtained in \cite{andrea}. The inside of the lake is ruled out by the bootstrap. The S-matrices being used in the main text, lie on the boundary of the allowed region.

\item River: This region is obtained on imposing sign restrictions on the D- and S-wave scattering lengths. The D-wave sign restrictions follow from dispersion relation considerations while the S-wave restrictions are observed in chiral perturbation theory as well as experimental results \cite{ABPHAS}. The allowed region looks like a river and has an upper and lower boundary. The S-matrices being used lie on these boundaries. Chiral perturbation theory lives at a kink like feature near the upper boundary \cite{ABPHAS}.

\end{enumerate}

The blue zones in fig. 1 in the main text are obtained by spanning over S-matrices allowed by the bootstrap considerations described above. In practice, we have 99 S matrices at our disposal to generate the plots.

\section{${\mathcal A}^{(2,3)}(t)$ and the S-matrix bootstrap}

In this section, we will investigate the question: How close are individual S-matrices in the S-matrix bootstrap to the Alexander polynomial ${\mathcal A}^{(2,3)}(t)=t+1/t-1=2\xi-1$. To quantify this, we will define the distance between two polynomials in the $\xi$ variable as:
\be
d(p_1|p_2)=\int_{-1}^1 d\xi (p_1(\xi)-p_2(\xi))^2\,.
\ee
Using this we can compare how close $\rho_1$ is to the Alexander polynomial for the pion S-matrices obtained from the bootstrap. From the pion bootstrap, we express $\rho_1$ as a function of $\xi$ where $\xi=\xi_{min}=1-\frac{243 a^2}{128}+\frac{729 a^3}{1024}$. Then we expand up to $O(\xi^6)$.  Focusing on the ``upper river'' we obtain the plot in fig.(\ref{fig:near}). As is evident from the figure, most S-matrices are far from the knot polynomial. Interestingly, the S-matrix which maximizes $\rho_1$ at $a=16/9$ is also the one which is nearest to the knot polynomial. All the S-matrices which have low values of $d(p_1|p_2)$ as exhibited in fig.(\ref{fig:near})(b)  are linear in $\xi$ and have slopes close to $2$ or $1$. For future work, it will be interesting to explore the connection between minimization of $d(p_1|p_2)$ and emerging integral slope in more detail.

\begin{figure}[hbt]
	\begin{subfigure}[b]{0.4\textwidth}
		\hskip 1.5cm
		\includegraphics[width=\textwidth]{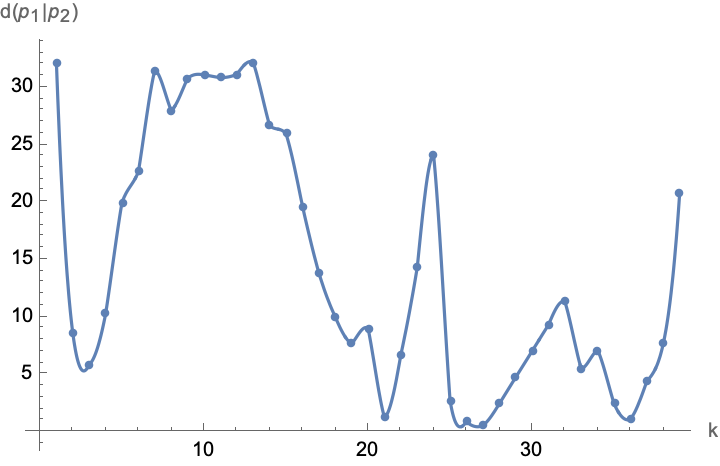}
		\caption*{\hskip 1cm (a) }
		\label{fig:near1}
	\end{subfigure}
	\hfill
	\begin{subfigure}[b]{0.4\textwidth}
		\hskip -1.5cm
		\includegraphics[width=\textwidth]{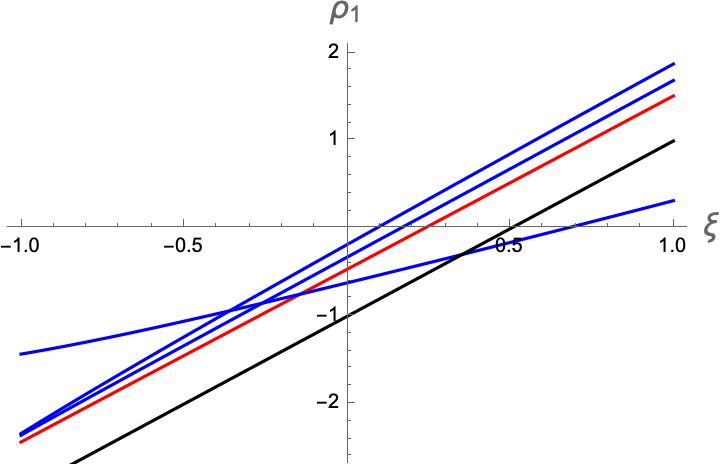}
		 \caption*{\hskip -3cm (b) }
		\label{fig:near2}
	\end{subfigure}
	\caption{(a) $d(p_1|p_2)$ vs $k$ where $k$ labels the S-matrix. (b) The $\rho_1$ vs $\xi$ for the $k=21,26,27,36$ S-matrices which have low values of $d(p_1|p_2)$. The $k=27$ which has the minimum value is indicated in red while the $\phi^3$ theory is in black.}
	\label{fig:near}
\end{figure}

\section{\bf Type II string theory and knot polynomials}
In this section, we will examine the type II tree level string amplitude for 2-2 dilaton scattering and show that the low energy expansion can be well approximated by a sum of knot polynomials. First we will recast the discussion about transcendentality in the amplitude in terms of the $\tilde z, a$ variables to make a statement about transcendentality and the knot polynomials. We will follow the discussions in \cite{greenwen, dhokergreen}.  
\subsection{Transcendentality and knots}
The amplitude, up to an overall kinematic factor of $(s_1 s_2+s_1 s_3+s_2 s_3)^2$ is given by
\begin{equation}\label{2exp}
\tilde A(s_1,s_2,s_3)=\frac{1}{s_1 s_2 s_3}\frac{\Gamma(1-s_1)\Gamma(1-s_2)\Gamma(1-s_3)}{\Gamma(1+s_1)\Gamma(1+s_2)\Gamma(1+s_3)}\,,
\end{equation}
where $s_1+s_2+s_3=0$. In terms of the $\tilde z, a$ variables introduced in the main text, we have the following expansion around $\tz=0$
\begin{eqnarray}
&&\tilde A-\frac{1}{s_1 s_2 s_3}-2\zeta(3)=54 a^2 \tilde z \left(-\zeta(5)+a\zeta(3)^2\right)\nonumber\\
&+&54 a^2 \tz^2 \left(-2\zeta(5)+2 a \zeta(3)^2+27 a^2 \zeta(7)+54 a^3 \zeta(3)\zeta(5)\nonumber \right. \\&+&\left. 9a^4 (\zeta(9)+2\zeta(3)^3)\right)\nonumber \\ &+&O(\tz^3)\,.
\end{eqnarray}
Following \cite{dhokergreen}, we assign the kinematic variable $a$ a transcendentality weight $-1$ and $\tz$ a weight $0$. $\zeta(n)$ has weight $n$. Then it is easy to see that each term in the above expansion has weight $3$. We can discuss this a bit more generally using the CSDR in eq.12 in the main text. First let us focus on the powers of $a$ that appear at a particular order in $\tz$. $U_n(\xi)$ is a degree $n$ polynomial while $\xi$ itself involves degree-2 and degree-3 terms in $a$. The measure $d\xi$ gives a factor $27 a^2(3a-2s_1')$ where we integrate over $s_1'$. Then, for instance at $\tz^2$ order, we have $n=1$ and hence the maximum degree of $a$ is 6. In a local theory this maximum degree cannot change. Specifically, we have all integer powers of $a$ from 2 to 6. This is precisely the pattern above. Then since we have $U_n(\xi)=\sum_{k=0}^n {\mathcal A}^{(2,2k+1)}(t)$ with $t+1/t=2\xi$, we easily note that the highest power of $a$ for a given $n$ is associated with the top knot polynomial ${\mathcal A}^{(2,2n+1)}(t)$. Together with the measure factor, the highest power of $a$ is $3n+3$. With the transcendentality assignments for $a$, we conclude that at a given order in $\tz^n$, ${\mathcal A}^{(2,2n+1)}(t)$ is associated with a maximum transcendentality weight of $3n$. Such relations between transcendentality and knot polynomials have been conjectured before in perturbative QFT--see \cite{kreimer}-- although the details differ. For instance in \cite{kreimer}, for perturbative $\phi^4$ theory, a correspondence between $(2,2n+1)$-torus knots and $\zeta(2n+1)$ has been proposed as opposed to $\zeta(3n)$ we find above. For $n=1$ they are the same but not otherwise.
\subsection{Low energy expansion in terms of knot polynomials}
Now let us examine numerically the expansion in eq.(\ref{2exp}). We rewrite the rhs (up to 2 decimal places) as
\begin{eqnarray}\label{numap}
&&\tz (-55.99 a^2+78.03 a^3)+\tz^2 (-111.99 a^2+156.05 a^3+1470.17 a^4\nonumber \\&&-3634.64 a^5+2175.24 a^6)+O(\tz^3)\,.
\end{eqnarray}
\begin{figure}[htb]
	\begin{subfigure}[b]{0.4\textwidth}
		\hskip 1.5cm
		\includegraphics[width=\textwidth]{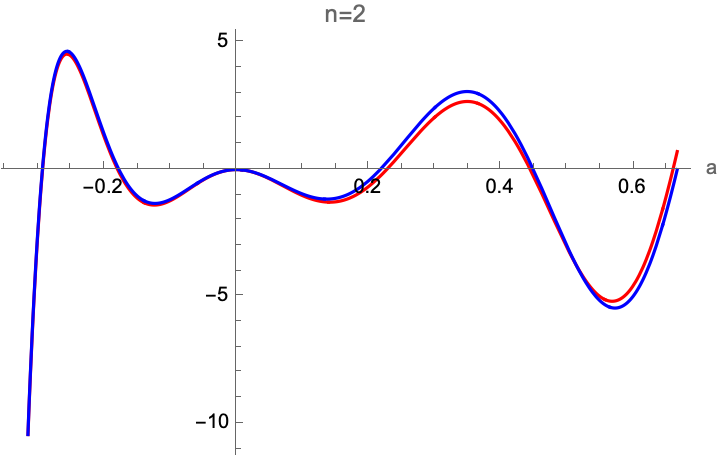}
		\caption*{\hskip 1cm (a) }
		\label{fig:n1st}
	\end{subfigure}
	\hfill
	\begin{subfigure}[b]{0.4\textwidth}
		\hskip -1.5cm
		\includegraphics[width=\textwidth]{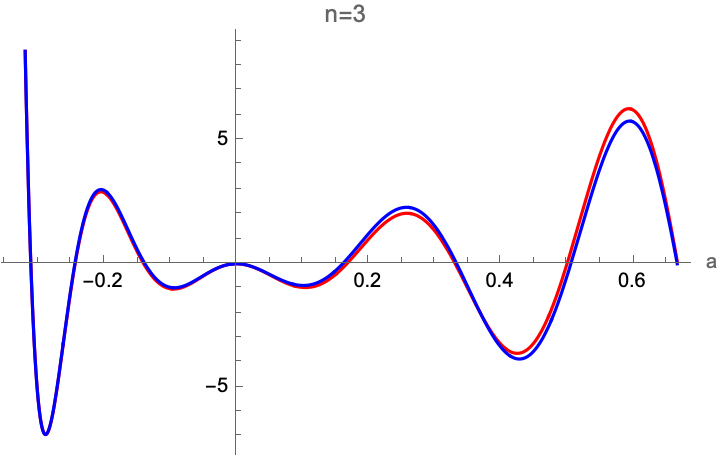}
		 \caption*{\hskip -3cm (b) }
		\label{fig:n2st}
	\end{subfigure}
	\caption{Comparison of the $\tz^{n+1}$ coefficients for string (red) and knot polynomial sum (blue) for $n=2,3$.}
	\label{fig:n1n2str}
\end{figure}

It can be expected that the low energy expansion should be dominated by the first massive pole (see appendix of \cite{ASAZ} for an explicit check). So in the dispersive integral should be well approximated by the lower limit of the integral which in our normalization starts at $1$. Then writing the measure factor's contribution as $\mu(a)\equiv 27 a^2(3a-2)$, our expectation is that eq.(\ref{numap}) is going to be approximated by (${\mathcal A}^{(2,1)}(t)=1$)
\begin{eqnarray}
&&\mu(a)\left(\tz {\mathcal A}^{(2,1)}+\tz^2({\mathcal A}^{(2,1)}+{\mathcal A}^{(2,3)})\right)+O(\tz^3)\,,\\
&=&\tz(-54 a^2+81 a^3)+\tz^2 (-108 a^2+162 a^3+1458 a^4\nonumber \\ &&-3645 a^5+2187 a^6)+O(\tz^3)\,. \label{strapp}
\end{eqnarray}
The agreement with eq.(\ref{numap}) is good! In fact for higher orders in $\tz$, the agreement becomes better as shown in fig.(\ref{fig:n1n2str}). This works mainly due to the fact that there is spin-0 dominance \cite{nima2, bern, SGPRAS} and hence the contribution from the Legendre polynomial in the CSDR is trivial.

\section{String amplitude in terms of $q$-oscillator}
Here I quote the formula for the type II tree level string amplitude, used in the main text, in terms of the $q$-oscillator representation. Explicitly, one can derive the fully crossing symmetric expansion
\be
\tilde A-\frac{1}{s_1 s_2 s_3}-2\zeta(3)=-{\mathcal{Z}}\sum_{k=1}^\infty \mu_k \langle b^\dagger b\rangle_{\beta, q_k}\,,
\ee
where ${\mathcal Z}$ and $\beta$ are as defined in the main text, and $q_k$ is defined via $2\xi=2-27(a/k)^2+27(a/k)^3=q_k+1/q_k$ and
\be
\mu_k=\frac{27 a^2 (\frac{3a}{k}-2)}{ k^3 (k!)^2} (-1)^k \left(1-\frac{k}{2}(1-\lambda_k)\right)_{k-1} \left(1-\frac{k}{2}(1+\lambda_k)\right)_{k-1}  \,,
\ee
with $\lambda_k=\sqrt{\frac{k+3a}{k-a}}$ and $(a)_b$ being the Pochhammer symbol, making it clear that $\mu_k$ is a degree $k-1$ polynomial in $(\lambda_k)^2$. Between $-1/3\leq a \leq 2/3$, which is where $q$ is complex with $|q|=1$, it can be checked that $\mu_k\geq 0$. Note that there is a quantization in the $q$-deformation in terms of the level $k$. For $k\rightarrow \infty$, $q_k\rightarrow 1$ while $\mu_k\rightarrow 54 \frac{k^{2a-5}}{\Gamma^2[a]}$. When $\tz\rightarrow 0$, $k=1$ dominates \cite{foot7}. The locality constraints \cite{ASAZ} are equivalent to the statement that $\sum_{k=1}^\infty [n]_{q_k}\mu_k=P_{3n}(a)$ where $P_k(a)$ is a degree-$k$ polynomial in $a$. Put differently, for a given $n$, the top knot polynomial in the description is the torus-$(2,2n+1)$. The simplicity of these conditions should enable a systematic study.


\begin{thebibliography}{99}
\bibitem{ffoot} While this has not been explicitly pointed out in the literature, it is easy to verify that it is true.
\bibitem{foot1}{We could have also considered the massive $\phi^3$ theory in which case we can just work with the shifted Mandelstam invariants $s=s_1+4/3 m^2$ etc. so that $s_1+s_2+s_3=0$.}
\bibitem{foot2}{Using eq.(\ref{torusA}), eq.(\ref{simpgen}) can be checked to be $\frac{\beta}{m^2} \frac{\tz}{(\tz-t)(\tz-\frac{1}{t})}$ as expected.}
\bibitem{ffoot3} A function $f(z)$ is typically real in a domain, if $Im\,z Im\, f(z)$ has the same sign in that domain. These functions are also called Herglotz. See \cite{RS} for more details.
\bibitem{foot3}{In \cite{spinning}, such CSDRs have been generalized for external particles carrying spin, in the context of weakly coupled EFTs.}
\bibitem{pionsupp1} Further details about how the pion S-matrices were obtained can be found in the appendix.
\bibitem{foot6}{$f(z)$ is typically real if $Im f(z) Im z \geq 0$ when $Im z\neq 0$. If  $Im f(z) Im z \leq 0$, we use $-f(z)$. See \cite{RS} for a review of the relevant mathematics.}
\end{thebibliography}
\end{document}